\documentclass{PoS-hep}
\usepackage{amsmath}
\usepackage{bm}
\usepackage{epsfig}
\usepackage{graphics}
\usepackage{upgreek}
\usepackage{amsfonts}
\usepackage{amsbsy}
\usepackage{amscd}
\usepackage{bbm}
\usepackage{eufrak}
\usepackage{cite}

\renewenvironment{subequations}{%
\refstepcounter{equation}%
\setcounter{parentequation}{\value{equation}}%
  \setcounter{equation}{0}
  \ignorespaces
}{%
  \setcounter{equation}{\value{parentequation}}%
  \ignorespacesafterend
}

\newcommand{\beqs}{\begin{subequations}}
\newcommand{\eeqs}{\end{subequations}}
\newcommand{\eec}{\end{center}}
\newcommand{\bec}{\begin{center}}
\newcommand{\eem}{\end{matrix}}
\newcommand{\bem}{\begin{matrix}}
\newcommand{\Eref}[1]{Eq.~(\ref{#1})}
\newcommand{\Sref}[1]{Sec.~\ref{#1}}
\newcommand{\Fref}[1]{Fig.~\ref{#1}}
\newcommand{\Tref}[1]{Table~\ref{#1}}
\newcommand{\cref}[1]{Ref.~\cite{#1}}

\newcommand\eqs[2]{Eqs.~(\ref{#1}) and (\ref{#2})}

\newcommand{\fr}{\ensuremath{f_{1}}}
\newcommand{\frr}{\ensuremath{f_{2}}}
\newcommand{\fk}{\ensuremath{f_p}}
\newcommand{\fkk}{\ensuremath{N_p}}

\newcommand{\ks}{\ensuremath{k_\star}}

\newcommand{\eeq}{\end{equation}}
\newcommand{\beq}{\begin{equation}}
\newcommand{\ba}{\begin{array}}
\newcommand{\ea}{\end{array}}
\newcommand{\bea}{\begin{eqnarray}}
\newcommand{\eea}{\end{eqnarray}}
\newcommand{\ftn}{\footnotesize}

\newcommand{\ssz}{\scriptsize}

\newcommand{\tr}{{\mbox{\sf\ssz T}}}

\renewcommand{\Re}{{\mbox{\sf\small Re}}}

\newcommand{\hepph}[1]{{\ftn \tt hep-ph/#1}}

\newcommand{\arxiv}[1]{{\ftn\tt  arXiv:#1}}

\newcommand\vev[1]{\langle {#1} \rangle}
\newcommand\vevi[1]{\langle {#1} \rangle_{\rm I}}
\def\lf{\left(}
\def\rg{\right)}

\def\llgm{\left\lgroup}
\def\rrgm{\right\rgroup}

\newcommand{\Vhi}{\ensuremath{V_{\rm I}}}
\newcommand{\Hhi}{\ensuremath{H_{\rm I}}}
\newcommand{\as}{\ensuremath{\alpha_{\rm s}}}
\newcommand{\wrh}{\ensuremath{w_{\rm rh}}}

\newcommand{\Dex}{\ensuremath{\Delta_{\star}}}

\newcommand{\Ve}{\ensuremath{V}}

\newcommand{\Qef}{\ensuremath{\Lambda_{\rm UV}}}

\newcommand{\plk}{{\slshape\small Planck}}

\def\bbet{{\bar\beta}}
\def\al{{\alpha}}
\def\bt{{\beta}}

\def\l{\lambda}

\newcommand{\Trh}{\ensuremath{T_{\rm rh}}}

\newcommand{\ld}{\ensuremath{\lambda}}

\newcommand{\kp}{\ensuremath{\kappa}}
\newcommand{\se}{\ensuremath{\widehat\phi}}
\newcommand{\sex}{\ensuremath{\widehat{\phi}_*}}
\newcommand{\sgx}{\ensuremath{\phi_\star}}
\newcommand{\lda}{\ensuremath{\lambda_1}}
\newcommand{\ldb}{\ensuremath{\lambda_2}}

\newcommand{\ldd}{\ensuremath{\lambda_4}}

\newcommand{\sef}{\ensuremath{\widehat{\phi}_{\rm f}}}

\newcommand{\eph}{\ensuremath{\widehat \epsilon}}
\newcommand{\ith}{\ensuremath{\widehat \eta}}
\newcommand{\mP}{\ensuremath{m_{\rm P}}}
\def\Ka{K\"{a}hler potential}

\newcommand{\sg}{\ensuremath{\phi}}

\renewcommand{\sigma}{\ensuremath{\phi}}

\newcommand{\sgf}{\ensuremath{\phi_{\rm f}}}

\newcommand{\GeV}{{\mbox{\rm GeV}}}
\newcommand{\what}{\ensuremath{\widehat}}
\newcommand{\Khi}{\ensuremath{K}}

\newcommand{\Vhio}{\ensuremath{V_{\rm I0}}}

\newcommand{\diag}{\mbox{\sf\small diag}}

\newcommand{\Mgut}{\ensuremath{M_{\rm GUT}}}
\newcommand{\Ggut}{\ensuremath{G_{\rm GUT}}}
\newcommand{\Gsm}{\ensuremath{G_{\rm SM}}}

\newcommand{\mbl}{\ensuremath{M_{BL}}}

\newcommand{\mgut}{\ensuremath{M_{\rm GUT}}}

\newcommand{\ldu}{\ensuremath{\uplambda}}
\newcommand{\ns}{\ensuremath{n_{\rm s}}}
\newcommand{\As}{\ensuremath{A_{\rm s}}}
\newcommand{\Ns}{\ensuremath{N_{\star}}}

\def\th{{\theta}}
\def\thb{{\bar\theta}}
\def\thn{{\theta_{\Phi}}}

\newcommand\mtt[4]{\mbox{
$\llgm\bem #1 &#2 \cr #3& #4\eem\rrgm$}}

\newcommand{\phc}{\ensuremath{\Phi}}
\newcommand{\phcb}{\ensuremath{\bar\Phi}}
\newcommand\mtta[4]{\mbox{
$\llgm\bem #1 &#2 \cr #3& #4\eem\rrgm$}}

\newcommand{\sgm}{\ensuremath{\phi_{\rm mx}}}
\newcommand{\sgn}{\ensuremath{\phi_{\rm mn}}}

\newcommand{\mma}{\ensuremath{M_1}}
\newcommand{\mmb}{\ensuremath{M_2}}
\newcommand{\mmc}{\ensuremath{M_4}}

\newcommand{\ts}{\ensuremath{\delta_{42}}}
\newcommand{\rs}{\ensuremath{\delta_{21}}}
\newcommand{\rss}{\ensuremath{r_{21}}}
\newcommand{\rrs}{\ensuremath{r_{12}}}
\newcommand{\tss}{\ensuremath{r_{42}}}
\newcommand{\tts}{\ensuremath{r_{24}}}

\renewcommand{\Dex}{\ensuremath{\Delta_{\rm \star}}}

\newcommand{\kbba}{\ensuremath{{K_{221}}}}

\newcommand{\tkbba}{\ensuremath{{\widetilde K_{221}}}}
\newcommand{\tkba}{\ensuremath{{\widetilde K_{21}}}}

\newcommand{\kas}{\ensuremath{K_{1\rm s}}}
\newcommand{\tkas}{\ensuremath{\widetilde K_{1\rm s}}}
\newcommand{\kbas}{\ensuremath{K_{21\rm s}}}
\newcommand{\tkbas}{\ensuremath{\widetilde K_{21\rm s}}}

\newcommand{\kb}{\ensuremath{K_{2}}}
\newcommand{\kba}{\ensuremath{K_{21}}}

\def\sub{subplanckian}

\newcommand{\dci}{{\small\sf $\delta$CI}}
\newcommand{\ca}{{\small\sf CI2}}
\newcommand{\cb}{{\small\sf CI4}}
\newcommand{\dhi}{{\small\sf $\delta$HI}}
\newcommand{\ha}{{\small\sf HI4}}
\newcommand{\hb}{{\small\sf HI8}}

\title{Pole Inflation in Supergravity}

\ShortTitle{Pole Inflation in SUGRA}

\author{\speaker{C. Pallis}\\
Laboratory of Physics, Faculty of Engineering, \\ Aristotle
University of Thessaloniki, \\ GR-541 24 Thessaloniki, GREECE \\
E-mail: \email{kpallis@gen.auth.gr}}

\abstract{We show how we can implement, within Supergravity,
chaotic inflation in the presence of a pole of order one or two in
the kinetic mixing of the inflaton sector. This pole arises due to
the selected logarithmic K\"ahler potentials $K$, which
parameterize hyperbolic manifolds with scalar curvature related to
the coefficient $(-N)<0$ of a logarithmic term. The associated
superpotential $W$ exhibits the same $R$ charge with the
inflaton-accompanying superfield and includes all the allowed
terms. The role of the inflaton can be played by a gauge singlet
or non-singlet superfield. Models with one logarithmic term in $K$
for the inflaton, require $N=2$, some tuning -- of the order of
$10^{-5}$ -- between the terms of $W$ and predict a
tensor-to-scalar ratio $r$ at the level of $0.001$. The tuning can
be totally eluded for more structured $K$'s, with $N$ values
increasing with $r$ and spectral index close or even equal to its
present central observational value.
}

\FullConference{Corfu Summer Institute 2021 "School and Workshops
on Elementary Particle Physics and Gravity"\\ 29 August - 9
October 2021\\ Corfu, Greece }

\begin{document}

\section{Introduction}

Among the many scenarios of inflation, the one which stands out in
terms of its simplicity, elegance and phenomenological success is
\emph{chaotic inflation} ({\sf\small CI}). Most notably, the
power-law potentials, employed in models of CI, have the forms
\beq V_{\rm I}=\ld^2\sg^n/n\>\>\> \mbox{or}\>\>\>V_{\rm
I}=\ld^2(\sg^{n/2}-M^2)^2/n\>\>\>\mbox{for}\>\>\>M\ll\mP=1,\label{vn}\eeq
which are very common in physics and so it is easy the
identification of the inflaton $\sg$ with a field already present
in the theory. E.g., within \emph{Higgs inflation} ({\sf\small
HI}) the inflaton could play, at the end of inflation, the role of
a Higgs field. However, for $n=2$ and $4$ the theoretically
derived values for spectral index $\ns$ and/or tensor-to-scalar
ratio $r$ are not consistent with the observational ones
\cite{plin}. A way out of these inconsistencies is to introduce
some non-minimality in the gravitational or the kinetic sector of
the theory. In this talk, which is based on
Refs.~\cite{sor,epole}, we focus on the latter possibility.
Namely, our proposal is tied to the introduction of a pole in the
kinetic term of the inflaton field. For this reason we call it for
short \emph{Pole} (chaotic) \emph{inflation} ({\sf\small PI})
\cite{terada}.

Below, we first briefly review the basic ingredients of PI in a
non-\emph{Supersymmetric} ({\sf\small SUSY}) framework
(Sec.~\ref{Fhi}) and constrain the parameters of two typical
models in Sec.~\ref{nmi} taking into account the observational
requirements described in \Sref{obs}. Throughout the text, the
subscript $,\chi$ denotes derivation \emph{with respect to}
({\sf\small w.r.t}) the field $\chi$, charge conjugation is
denoted by a star ($^*$) and we use units where the reduced Planck
scale $\mP = 2.44\cdot 10^{18}~\GeV$ is set equal to unity.

\subsection{Non-SUSY Set-up}\label{Fhi}

The lagrangian of the homogenous inflaton field $\sg=\sg(t)$ with
a kinetic mixing takes the form
\beq \label{action1} {\cal  L} = \sqrt{-\mathfrak{g}}
\left(\frac{\fkk}{2\fk^2} \dot\sg^2-
\Vhi(\sg)\right)~~\mbox{with}~~\fk=1-\sg^p,~p>0~~\mbox{and}~~\fkk>0.\eeq
Also, $\mathfrak{g}$ is the determinant of the background
Friedmann-Robertson-Walker metric $g^{\mu\nu}$ with signature
$(+,-,-,-)$ and dot stands for derivation w.r.t the cosmic time.
Concentrating on integer $p$ values we can derive the canonically
normalized field, $\se$, as follows
\beq \label{VJe}
\frac{d\se}{d\phi}=J=\frac{\sqrt{\fkk}}{\fk}~~\Rightarrow~~\se=\frac{\sqrt{\fkk}}{p}B(\sg^p;1/p,0),
\eeq
where $B(z;m,l)$ represents the incomplete Beta function. Note
that $\se$ gets increased above unity for $p<10$ and
$0\leq\sg\lesssim1$, facilitating, thereby, the attainment of PI
with \sub\ $\sg$ values. Inverting this function we obtain, e.g.,
\beq \label{sesg} \sg= \begin{cases}
1-e^{-\se/\sqrt{N_1}} &\mbox{for}~~p=1,\\
\tanh{\lf\frac{\se}{\sqrt{N_2}}\rg} &\mbox{for}~~p=2\,.
\end{cases}\eeq
%
As a consequence, \Eref{action1} can be brought into the form
\beq {\cal  L} = \sqrt{-\mathfrak{g}} \left(
\frac12\dot\se^2-\Vhi(\sg(\se))\right). \label{action} \eeq
For $\se\gg1$, $V_{\rm I}(\se)$ -- expressed in terms of $\se$ --
develops a plateau, and so it becomes suitable for driving
inflation of chaotic type called \emph{E-Model Inflation}
\cite{alinde, linde21} (or $\alpha$-Starobinsky model
\cite{ellis21}) and \emph{T-Model Inflation} \cite{tmodel,
linde21} for $p=1$ and $2$ respectively.

\subsection{Inflationary Observables -- Constraints} \label{obs}

The analysis of PI can be performed using the standard slow-roll
approximation as analyzed below, together with the relevant
observational and theoretical requirements that should be imposed.

\subparagraph{\bf (a)} The number of e-foldings $\Ns$ that the
scale $\ks=0.05/{\rm Mpc}$ experiences during PI must be enough
for the resolution of the  problems of standard Big Bang, i.e.,
\cite{plcp}
\begin{equation} \label{Nhi}  \Ns=\int_{\sef}^{\sex}
d\se\frac{\Vhi}{\Ve_{\rm I,\se}}\simeq61.3+\frac{1-3w_{\rm
rh}}{12(1+w_{\rm rh})}\ln\frac{\pi^2g_{\rm
rh*}\Trh^4}{30\Vhi(\sgf)}+\frac14\ln{\Vhi(\sgx)^2\over g_{\rm
rh*}^{1/3}\Vhi(\sgf)},\eeq
where $\sex$ is the value of $\se$ when $\ks$ crosses the
inflationary horizon whereas $\se_{\rm f}$ is the value of $\se$
at the end of PI, which can be found, in the slow-roll
approximation, from the condition
\beqs\beq\mbox{\sf\small max}\{\epsilon(\sg_{\rm
f}),|\eta(\sg_{\rm f})|\}=1,\>\>\>~\mbox{where}\>\>\>\epsilon=
{1\over2}\left(\frac{\Ve_{\rm I,\se}}{\Ve_{\rm
I}}\right)^2\>\>\>\mbox{and}\>\>\>\eta= \frac{\Ve_{\rm
I,\se\se}}{\Ve_{\rm I}}\,.\label{sr}\eeq
Also we assume that PI is followed in turn by an oscillatory phase
with mean equation-of-state parameter $w_{\rm rh}$, radiation and
matter domination. We determine it applying the formula
\cite{epole}
\beq w_{\rm rh}=2\frac{\int_{\sgn}^{\sgm} d\sg J(1-
\Vhi/\Vhi(\sgm))^{1/2}}{\int_{\sgn}^{\sgm} d\sg J(1-
\Vhi/\Vhi(\sgm))^{-1/2}}-1,\label{wrh}\eeq\eeqs
where $\sgn=\vev{\sg}$ is the \emph{vacuum expectation value}
({\small\sf v.e.v}) of $\sg$ after PI and $\sgm$ is the amplitude
of the $\sg$ oscillations \cite{epole}. Motivated by
implementations \cite{univ} of non-thermal leptogenesis, which may
follow PI, we set $\Trh\simeq10^9~\GeV$ for the reheat
temperature. Indicative values for the energy-density effective
number of degrees of freedom include $g_{\rm rh*}=106.75$ or
$228.75$ corresponding to the \emph{Standard Model} ({\small\sf
SM}) or \emph{Minimal SUSY SM} ({\small\sf MSSM}) spectrum
respectively.

\subparagraph{\bf (b)}  The amplitude $\As$ of the power spectrum
of the curvature perturbations generated by $\sg$ at  $\ks$ has to
be consistent with data~\cite{plcp}, i.e.,
\begin{equation}  \label{Prob}
A_{\rm s}={\Ve_{\rm I}(\sex)^{3}}/{12\, \pi^2}{\Ve_{\rm
I,\se}(\sex)^2} \simeq2.105\cdot 10^{-9}\,.
\end{equation}

\subparagraph{\bf (c)} The remaining inflationary observables
($\ns$, its running $\as$ and $r$) have to be consistent with the
latest \emph{Planck release 4} ({\sf\small PR4}), \emph{Baryon
Acoustic Oscillations} ({\sf\small BAO}), CMB-lensing and
BICEP/{\it Keck} ({\sf\small BK18}) data \cite{plin,gws}, i.e.,
\begin{equation}  \label{nswmap}
\mbox{\sf
(i)}\>\>\ns=0.965\pm0.009\>\>\>~\mbox{and}\>\>\>~\mbox{\sf
(ii)}\>\>r\leq0.032,
\end{equation}
at 95$\%$ \emph{confidence level} ({\sf c.l.}) -- pertaining to
the $\Lambda$CDM$+r$ framework with $|\as|\ll0.01$. These
observables are estimated through the relations
\beq\label{ns} \mbox{\sf (i)}\>\>\ns=\: 1-6\eph_\star\ +\
2\ith_\star,\>\>\>\mbox{\sf (ii)}\>\> \as
=\frac23\left(4\ith^2-(\ns-1)^2\right)-2\what\xi_\star\>\>\>~
\mbox{and}\>\>\>~\mbox{\sf (iii)}\>\>r=16\eph_\star\,, \eeq
with $\xi={\Ve_{\rm I,\se} \Ve_{\rm I,\se\se\se}/\Ve_{\rm I}^2}$
-- the variables with subscript $\star$ are evaluated at
$\sg=\sgx$.

\subparagraph{\bf (d)}  The effective theory describing PI has to
remain valid up to a UV cutoff scale $\Qef\simeq\mP$ to ensure the
stability of our inflationary solutions, i.e.,
\beq \label{uv}\mbox{\sf (i)}\>\> \Vhi(\sgx)^{1/4}\leq\Qef
\>\>\>~\mbox{and}\>\>\>~\mbox{\sf (ii)}\>\>\sgx\leq\Qef.\eeq

\subsection{Results}\label{nmi}

Using the criteria of \Sref{obs}, we can now analyze the
inflationary models based on the potential in \Eref{vn} and the
kinetic mixing in \Eref{action1} for $p=1$ and $2$. The slow-roll
parameters are
\beq\epsilon=\frac{n^2\fk}{2\fkk\sg^2}\>\>\mbox{and}\>\>\eta=\frac{n\fk}{\fkk\sg^2}\lf
n-1-(n+p-1)\sg^p\rg,\label{sr0}\eeq
whereas from \Eref{Nhi} we can compute
\beq \label{Ns0} \Ns\simeq\begin{cases}N_1\lf \sgx+f_{1\star}\ln
f_{1\star}\rg/n f_{1\star} & \mbox{for}~~p=1,\\  N_2\sgx^2/2n
f_{2\star} &\mbox{for}~~p=2, \end{cases}\eeq
where $f_{p\star}=f_p(\sgx)$. Since $f_{p\star}$ appears in the
denominator, $\Ns$ increases drastically as $\sgx$ approaches
unity, assuring thereby the achievement of efficient PI. The
relevant tuning can be somehow quantified defining the quantity
\beq \Dex=1 - \sgx.\label{dex}\eeq
The naturalness of the attainment of PI increases with $\Dex$.
Imposing the condition of \Eref{sr} and solving \Eref{Ns0} w.r.t
$\sgx$, we arrive at
\beq \label{sgx0} \sgf\ll\sgx\simeq\begin{cases} n\Ns/(n\Ns+N_1)& \mbox{for}~~p=1,\\
\sqrt{2n\Ns/(2n\Ns+N_2)}&\mbox{for}~~p=2, \end{cases}\eeq
where we neglect the logarithmic contribution in the first of the
relations in \Eref{Ns0}. We remark that PI is attained for $\sg<1$
-- and so \Eref{uv} is fulfilled -- thanks to the location of the
pole at $\sg=1$. On the other hand, \Eref{Prob} implies
\beq \label{lan0} \ld\simeq\lf{\sqrt{3nN\As}\pi/\Ns}\rg\begin{cases} 2& \mbox{for}~~p=1,\\
1&\mbox{for}~~p=2\,. \end{cases}\eeq
From \Eref{ns} we obtain the model's predictions, i.e.,
\beq \label{ns0} \ns\simeq1-{2/\Ns}, \>\>\> \as
\simeq-{2/N_\star^2}\>\>\>
\mbox{and}\>\>\> r\simeq\begin{cases}8N_1/\Ns^2& \mbox{for}~~p=1,\\
2N_2/\Ns^2 &\mbox{for}~~p=2\,,\end{cases} \eeq
which are independent of $n$ and for this reason these models are
called $N$-attractors \cite{tmodel, alinde, linde21, ellis21}.
However, the variation of $n$ in \Eref{vn} generates a variation
to $\wrh$ in \Eref{wrh} and via \Eref{Nhi} to $\Ns$ which slightly
distinguishes the predictions above. E.g., fixing $N_1=10$ we
obtain
\beqs\beq \wrh\simeq\begin{cases}-0.08,\\
~~~0.19,\end{cases}\Ns\simeq\begin{cases}49.4,\\
54.6,\end{cases}\Dex\simeq\begin{cases}0.074,\\
0.04,\end{cases}\ns\simeq\begin{cases}0.963\\
0.965\end{cases}r\simeq0.02 ~~\mbox{for}~~~ n=\begin{cases}2,\\
4\end{cases}\label{nsn1}\eeq
and $p=1$ with $\as\sim 10^{-4}$. Similar $\as$ values are
obtained setting $N_2=10$ and $p=2$ which yields
\beq \wrh\simeq\begin{cases}-0.04,\\
~~~0.23,\end{cases}\Ns\simeq\begin{cases}50.2,\\
54.6,\end{cases}\Dex\simeq\begin{cases}0.024,\\
0.01,\end{cases} \ns\simeq\begin{cases}0.962,\\
0.963,\end{cases}r\simeq\begin{cases}0.0074,\\
0.0064,\end{cases}\mbox{for}~~n=\begin{cases}2.\\
4.\end{cases}\label{nsn2}\eeq\eeqs  Notice that $\Dex$ is larger
for $p=1$. Imposing the bound on $r$ in \Eref{nswmap}, we can find
a robust upper bound on $N_p$. Namely, we find numerically
\beq N_1\lesssim 19 ~~\mbox{and}~~ N_2\lesssim 55.
\label{N12b}\eeq

Therefore, we can conclude that the presence of $\fk$ in
\Eref{action1} revitalizes CI rendering it fully consistent with
the present data in \Eref{nswmap} without introducing any
complication with the validity of the effective theory. Recall
\cite{corfu12} that the last problem plagues models of CI with
large non-minimal coupling to gravity for $n>2$.

\subsection{Outline}\label{plan}

It would be certainly interesting to inquire if it is possible to
realize  similar models of PI in a SUSY framework where a lot of
the problems of SM are addressed. We below describe how we can
formulate PI in the context of \emph{Supergravity} ({\sf\small
SUGRA}) in \Sref{fhim} and we specify six models of PI: three
models (\dci, \ca, \cb) employing a gauge singlet inflaton in
\Sref{fhi3} and three (\dhi, \ha, \hb) with a gauge non-singlet
inflaton in \Sref{fhi1}.

\section{Realization of PI Within SUGRA}\label{fhim}

We start our investigation presenting the basic formulation of
scalar theory within SUGRA in \Sref{sugra1} and then -- in
\Sref{sugra2} -- we outline our strategy in constructing viable
models of PI.

\subsection{General Set-up} \label{sugra1}

The part of the SUGRA lagrangian including the (complex) scalar
fields $Z^\al$ can be written as
\beqs \beq\label{Saction1} {\cal  L} = \sqrt{-\mathfrak{g}} \lf
K_{\al\bbet} D_\mu Z^\al D^\mu Z^{*\bbet}-V_{\rm SUGRA}\rg, \eeq
where the kinetic mixing is controlled by the K\"ahler potential
$K$ and the relevant metric defined as
\beq \label{kddef} K_{\al\bbet}={\Khi_{,Z^\al
Z^{*\bbet}}}>0\>\>\>\mbox{with}\>\>\>K^{\bbet\al}K_{\al\bar
\gamma}=\delta^\bbet_{\bar \gamma}.\eeq
Also, the covariant derivatives for the scalar fields $Z^\al$ are
given by
\beq D_\mu Z^\al=\partial_\mu Z^\al+ig A^{\rm a}_\mu T^{\rm
a}_{\al\bt} Z^\bt\eeq
with $A^{\rm a}_\mu$ being the vector gauge fields, $g$ the
(unified) gauge coupling constant and $T^{\rm a}$ with ${\rm
a}=1,...,\mbox{\sf\small dim}\Ggut$ the generators of a gauge
group $\Ggut$. Here and henceforth, the scalar components of the
various superfields are denoted by the same superfield symbol.

The SUGRA scalar potential, $V_{\rm SUGRA}$, is given in terms of
$K$, and the superpotential, $W$, by
\beq V_{\rm SUGRA}=\Ve_{\rm F}+ \Ve_{\rm D}\>\>\>\mbox{with}\>\>\>
\Ve_{\rm F}=e^{\Khi}\left(K^{\al\bbet}{\rm F}_\al {\rm
F}^*_\bbet-3{\vert W\vert^2}\right) \>\>\>\mbox{and}\>\>\>\Ve_{\rm
D}= g^2 \sum_{\rm a} {\rm D}_{\rm a} {\rm D}_{\rm a}/2,
\label{Vsugra} \eeq
where a trivial gauge kinetic function is adopted whereas the F-
and D-terms read
\beq \label{Kinv} {\rm F}_\al=W_{,Z^\al}
+K_{,Z^\al}W\>\>\>\mbox{and}\>\>\>{\rm D}_{\rm a}= Z_\al\lf T_{\rm
a}\rg^\al_\bt K^\bt\>\>\>\mbox{with}\>\>\>
K^{\al}={\Khi_{,Z^\al}}\,.\eeq\eeqs
Therefore, the models of PI in \Sref{Fhi} can be supersymmetrized,
if we select conveniently the functions $K$ and $W$ so that
\eqs{vn}{action1} are reproduced.

\subsection{Modeling PI in SUGRA}\label{sugra2}

We concentrate on PI driven by $V_{\rm F}$. To achieve this, we
have to assure that $V_{\rm D}=0$ during PI. This condition may be
attained in the following two cases:

\begin{itemize}
\item If the inflaton is (the radial part of) a gauge singlet
superfield $Z^2:=\Phi$. In this case, $\Phi$ has obviously zero
contribution to $V_{\rm D}$.

\item If the inflaton is the radial part of a conjugate pair of
Higgs superfields, $Z^2:=\Phi$ and $Z^3:=\bar\Phi$, which are
parameterized so as $V_{\rm D}=0$ -- see \Sref{fhi1}.
\end{itemize}

To achieve a kinetic term in \Eref{Saction1} similar to that in
\Eref{action1} for $p=1$ and $2$, we need to establish suitable
$K$'s so that
\beq \vevi{K}= -N
\ln\fk~~~\mbox{and}~~~\vevi{K_{\al\bbet}}=N/\fk^2\eeq
with $N$ related to $N_p$ -- here and henceforth the symbol
``$\vevi{Q}$" denotes the value of a quantity $Q$ during PI.
However, from the F-term contribution to \Eref{Vsugra}, we remark
that $K$ affects besides the kinetic mixing $V_{\rm SUGRA}$,
which, in turn, depends on the $W$ too. Therefore, $\fk$ is
generically expected to emerge also in the denominator of $V_{\rm
SUGRA}$ making difficult the establishment of an inflationary era.
This problem can be surpassed \cite{sor, epole} by two alternative
strategies:

\begin{itemize}

\item Adjusting $W$ and constraining the prefactor of $K$'s, so
that the pole is removed from $V_{\rm SUGRA}$ thanks to
cancellations \cite{sor,epole,eno5} which introduce some tuning,
though.

\item Adopting a structured $K$ which yields the desired kinetic
terms in \Eref{action1} but remains invisible from $V_{\rm SUGRA}$
\cite{tkref, sor, epole}. In a such case, any tuning on the $W$
parameters can be eluded.

\end{itemize}

In Sec.~\ref{fhi3} and \ref{fhi1} we show details on the
realization of these scenaria, taking into account that $f_1$ in
\Eref{action1} can be exclusively associated with a gauge singlet
inflaton whereas $f_2$ can be related to a gauge non-singlet
inflaton.

We reserved $\al=1$ for a gauge singlet superfield, $Z^1=S$ called
stabilizer or goldstino, which assists \cite{rube} us to formulate
PI of chaotic type in SUGRA. Its presence in $W$ is determined as
follows:

\begin{itemize}

\item It appears linearly in $W$ multiplying its other terms. To
achieve technically such a adjustment, we require that $S$ and $W$
are equally charged under a global $R$ symmetry.

\item It generates for $\vevi{S}=0$ the inflationary potential via
the only term of $V_{\rm SUGRA}$ in \Eref{Vsugra} which remains
alive
\beq \Vhi=\vevi{V_{\rm F}}= \vevi{e^{K}K^{SS^*}|W_{,S}|^2}.
\label{Vhio}\eeq
\item It assures the boundedness of $\Vhi$. Indeed, if we set
$\vevi{S}=0$, then $\vevi{K_{,z^\al}W}=0$ for $\al\neq1$ and
$-3\vert \vevi{W}\vert^2=0$. Obviously, non-vanishing values of
the latter term may render $\Ve_{\rm F}$ unbounded from below.
\item It can be stabilized at $\vevi{S}=0$ without invoking higher
order terms, if we select \cite{su11}
\beq \label{K2}
K_2=N_S\ln\lf1+|S|^2/N_S\rg~~\Rightarrow~~\vevi{K_2^{SS^*}}=1~~\mbox{with
$0<N_S<6$}.\eeq $K_2$ parameterizes the compact manifold
$SU(2)/U(1)$. Note that for $\vevi{S}=0$, $S$ is canonically
normalized and so we do not mention it again henceforth.

\end{itemize}

\section{PI With a Gauge Singlet Inflaton}\label{fhi3}

The SUGRA setup for this case is presented in \Sref{fhi30} and
then -- in \Sref{fhi31} -- we describe the salient features of
this model and we expose our results in \Sref{num3}.

\subsection{SUGRA Set-up}\label{fhi30}

This setting is realized in presence of two gauge singlet
superfields $S$ and $\Phi$. We adopt the most general
renormalizable $W$ consistent with the $R$ symmetry mentioned in
\Sref{sugra2}, i.e.,
\beq W= S\lf \lda \phc+\ldb\phc^2-M^2\rg \label{whi} \eeq
where $\lda, \ldb$ and $M$ are free parameters. As regards $K$,
this includes, besides $K_2$ in \Eref{K2}, one of the following
$K$'s, $\kas$ or $\tkas$, which yield a pole of order one in the
kinetic term of $\phc$ and share the same geometry -- see
\cref{epole}. Namely,
\beq \kas=-N\ln\left(1-(\phc+\phc^*)/2\right)~~~\mbox{or}~~~
\tkas=-N\ln\frac{(1-\phc/2-\phc^*/2)}{(1-\phc)^{1/2}(1-\phc^*)^{1/2}},\label{tkas}\eeq
with $\Re(\phc)<1$ and $N>0$. We opt a pole of order one as the
simplest choice, although models with a pole of order two were
also proposed \cite{alinde}. The $K$'s above are invariant under
the set of transformations composing a set of matrices which can
be related \cite{epole} to the group $U(1,1)$. Based on the $K$'s
above, we can define the following three versions of PI:

\begin{itemize}
\item \dci, where the total $K$ is chosen as
\beqs\beq \kbas=\kb+\kas. \label{kbas}\eeq
The elimination of pole in $\Vhi$ discussed above can be applied
if we set \beq N=2~~~\mbox{and}~~~\rss=-\ldb/\lda\simeq
1+\rs\,~~\mbox{with}~~~\rs\sim0~~~\mbox{and}~~M\ll1\label{kbas1}
\eeq\eeqs
such that the denominator including the pole in $\Vhi$ is (almost)
cancelled out.

\item \ca\ and \cb, which do not display any denominator in $\Vhi$
employing \beq \tkbas=\kb+\tkas \label{tkbas}\eeq with free
parameters $N$, $\lda$, $\ldb$ and $M$. The discrimination of
these models depends on which of the two inflaton-dependent terms
in \Eref{whi} dominates -- see below.
\end{itemize}

\subsection{Structure of the Inflationary Potential}\label{fhi31}

An inflationary potential of the type in \Eref{vn} can be derived
from \Eref{Vhio} specifying the inflationary trajectory as follows
\beq \vevi{S}=0\>\>\>\mbox{and}\>\>\>\vevi{\th}:=\mbox{\sf\small
arg}\vevi{\Phi}=0. \label{inftr3}\eeq
Inserting the quantities above into \Eref{Vhio} and taking into
account \Eref{K2} and
\beq \label{eK} \vevi{e^{K}}=\begin{cases}
\fr^{-N}&\mbox{for}\>\>\>K=\kbas,\\
1& \mbox{for}\>\>\>K=\tkbas,
\end{cases}\eeq
we arrive at the following master equation
\beq\Vhi=\ld^2 \begin{cases}
{\lf  \sg-\rss\sg^2-\mma^2\rg^2}/{\fr^{N}}&\mbox{for \dci},\\
{\lf \sg-\rss\sg^2-\mma^2\rg^2}&\mbox{for \ca},\\
{\lf \sg^2-\rrs\sg-\mmb^2\rg^2}&\mbox{for
\cb},\end{cases}\label{Vmab}\eeq
where $\sg=\Re(\Phi)$, $r_{ij}=-\ld_i/\ld_j$ with $i,j=1,2$ and
$\ld$ and $M_i$ are identified as follows
\beq\ld= \begin{cases}
\lda~~\mbox{and}~~\mma=M/\sqrt{\lda}&\mbox{for~\dci\ and \ca},\\
\ldb~~\mbox{and}~~\mmb=M/\sqrt{\ldb}&\mbox{for~\cb}.\end{cases}\label{ldab}\eeq
As advertised in \Sref{fhi30}, the pole in $\fr$ is presumably
present in $\Vhi$ of \dci, but it disappears for \ca\ and \cb. The
arrangement of \Eref{kbas1}, though, renders the pole harmless for
\dci.

The correct description of PI is feasible if we introduce the
canonically normalized fields, $\se$ and $\what \th$, as follows
\beq \label{K3} \vevi{K_{\Phi\Phi^*}}|\dot \Phi|^2
\simeq\frac12\lf\dot{\what\phi}^{2}+\dot{\what
\th}^{2}\rg~~\Rightarrow~~\frac{d\se}{d\sg}=J={\sqrt{N/2}\over\fr}~~\mbox{and}~~
\widehat{\theta}\simeq
J\sg\theta~~~\mbox{with}~~~\vevi{K_{\Phi\Phi^*}}=\frac{N}{4f_1^2}.
\eeq
We see that the relation between $\sg$ and $\se$ is identical with
\Eref{VJe} for $p=1$, if we do the replacement $N_1=N/2$. We
expect that \ca\ [\cb] yield similar results with the non-SUSY
models of PI with $p=1$ in \Eref{action1} and $n=2$ [$n=4$] in
\Eref{vn}, whereas \dci\ is totally autonomous.

\renewcommand{\arraystretch}{1.4}
\begin{table}[!t]
\bec\begin{tabular}{|c||c||c|l|l|}\hline {\sc Fields}&{\sc
Eigen-}& \multicolumn{3}{|c|}{\sc Masses Squared}\\\cline{3-5}
&{\sc states}& &
{$K=\kbas$}&{$K=\tkbas$}\\
\hline\hline
$1$ real scalar &$\widehat \theta$ & $\widehat m^2_{\theta}$& \multicolumn{2}{|c|}{$6H_{\rm I}^2$}\\
$2$ real scalars &$\what s_1,~\what s_2$ & $\what m^2_{
s}$&\multicolumn{2}{|c|}{$6H_{\rm I}^2/N_S$}\\\hline
$2$ Weyl spinors & ${(\what{\psi}_{\Phi}\pm
\what{\psi}_{S})/\sqrt{2}}$& $\what m^2_{
\psi\pm}$& \multicolumn{2}{|c|}{$6n(1-\sg)^2H_{\rm I}^2/N\sg^2$}\\
\hline
\end{tabular}\eec
\caption{\sl Mass spectrum of our CI models along the inflationary
trajectory of Eq.~(3.5) -- we take $n=1$ for \textsf{$\delta$\ftn
CI} and \textsf{\ftn CI2} whereas $n=2$ for \textsf{\ftn
CI4}.}\label{tab3}
\end{table}\renewcommand{\arraystretch}{1}

To check the stability of $V_{\rm SUGRA}$ in \Eref{Vsugra} along
the trajectory in \Eref{inftr3} w.r.t the fluctuations of
$Z^\alpha$'s, we construct the mass spectrum of the theory. Our
results are summarized in \Tref{tab3}. Taking into the limit
$\rs=\mma=0$ for \dci, $\rss=\mma=0$ for \ca\ and $\rrs=\mmb=0$
for \cb, we find the expressions of the masses squared $\what
m^2_{\chi^\al}$ (with $\chi^\al=\th$ and $s$) arranged in
\Tref{tab3}. We there display the masses $\what m^2_{\psi^\pm}$ of
the corresponding fermions too -- we define
$\what\psi_{\Phi}=J\psi_{\Phi}$ where $\psi_\Phi$ and $\psi_S$ are
the Weyl spinors associated with $S$ and $\Phi$ respectively. We
notice that the relevant expressions can take a unified form for
all models -- recall that we use $N=2$ in \dci\ -- and approach,
close to $\sg=\sgx\simeq1$, rather well the quite lengthy, exact
ones employed in our numerical computation. From them we can
appreciate the role of $N_S<6$ in retaining positive $\what
m^2_{s}$. Also, we confirm that $\what
m^2_{\chi^\al}\gg\Hhi^2\simeq\Vhio/3$ for $\sgf\leq\sg\leq\sgx$.

\begin{figure}[t]\vspace*{-0.6cm}
\begin{minipage}{75mm}
\includegraphics[height=9cm,angle=-90]{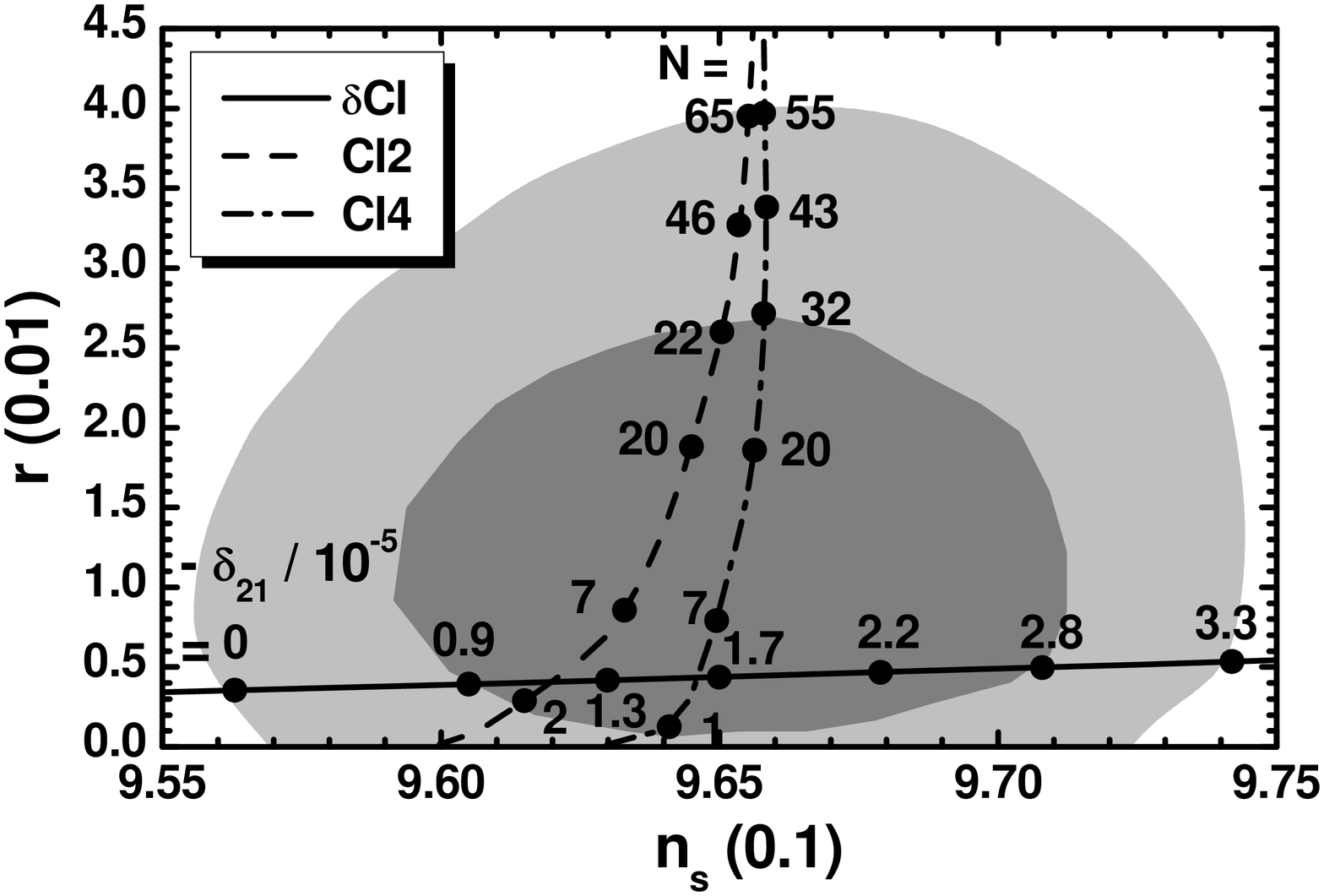}
\end{minipage}
\hfill
\begin{minipage}{95mm}
\begin{center}
{\small
\begin{tabular}{|c||c|c|c|}\hline
{Model:} &{\dci}&{\ca}&{\cb}\\\hline
${\rs}$ / $\rss$ / $\rrs$&$-1.7\cdot10^{-5}$&{$0.001$}&{$0.001$}\\
$N$&{$2$}&$10$&$10$\\\hline\hline
$\sgx/0.1$&$9.9$&$9.53$&{$9.84$}\\
$\Dex (\%)$&$1$&$4.7$&{$2$}\\
$\sgf/0.1$&{$6.66$}&$3.7$&{$5.6$}\\\hline
$\wrh$&$-0.24$&$-0.08$&{$0.26$}\\
$\Ns$&$44.4$&$51.5$&{$55.5$}\\\hline
$\ld/10^{-5}$&$1.2$&$2.1$&{$1.9$}\\\hline
$\ns/0.1$&$9.65$&$9.64$&{$9.65$}\\
$-\as/10^{-4}$&$11.4$ &$6.7$&{$6.2$}\\
$r/10^{-2}$&$0.44$&$1.3$&{$1.1$}\\\hline
\end{tabular}}
\end{center}
\end{minipage}
\caption{\sl Allowed curves in the $\ns-r$ plane for (i)
\textsf{\ftn $\delta$CI}, $\mma=0.01$ and various $\rs$'s
indicated on the solid line or (ii) \textsf{\ftn CI2}, $\mma=0.01$
and $\rss=0.001$ or \textsf{\ftn CI4}, $\mmb=0.01$ and
$\rrs=0.001$ and various $N$'s indicated on the dashed or
dot-dashed line respectively. The marginalized joint $68\%$
[$95\%$] c.l. regions \cite{gws} from PR4, {\sffamily\ftn BK18},
BAO and lensing data-sets are depicted by the dark [light] shaded
contours. The relevant field values, parameters and observables
corresponding to points shown in the plot are listed in the
Table.} \label{fig1}\end{figure}

\subsection{Results}\label{num3}

The dynamics of the analyzed models is analytically studied in
\cref{epole}. We here focus on the numerical results. After
imposing \eqs{Nhi}{Prob} the free parameters of
$$\mbox{\dci, ~\ca, ~\cb\ ~~are}~~~(\rs,\mma), (N, \rss, \mma)
~~\mbox{and}~~ (N,\rrs,\mmb), $$
respectively. Recall that we use $N=2$ exclusively for \dci.
Fixing $\mma=0.001$ for \dci,  $\mma=0.01$ and $\rss=0.001$ for
\ca\ and $\mmb=0.01$ and $\rrs=0.001$ for \cb, we obtain the
curves plotted and compared to the observational data in
\Fref{fig1}.  We observe that:

\subparagraph{\bf (a)} For \dci\ the resulting $\ns$ and $r$
increase with $|\rs|$ -- see solid line in \Fref{fig1}. This
increase, though, is more drastic for $\ns$ which covers the whole
allowed range in \Eref{nswmap}. From the considered data we
collect the results
\beq \label{resm1} 0\lesssim\rs/10^{-5}\lesssim3.3,
\>\>\>3.5\lesssim {r/10^{-3}}\lesssim5.3\>\>\>
\mbox{and}\>\>\>9\cdot10^{-3}\lesssim\Dex\lesssim0.01. \eeq
In all cases we obtain $\Ns\simeq44$ consistently with \Eref{Nhi}
and the resulting $\wrh\simeq-0.237$ from \Eref{wrh}. Fixing
$\ns=0.965$, we find $\rs=-1.7\cdot 10^{-5}$ and $r=0.0044$ -- see
the leftmost column of the Table in \Fref{fig1}.

\subparagraph{\bf (b)}  For \ca\ and \cb, $\ns$ and $r$ increase
with $N$ and $\Dex$ which increases w.r.t its value in \dci.
Namely, $\ns$ approaches its central observational value in
\Eref{nswmap} whereas the bound on $r$ yields an upper bound on
$N$.  More quantitatively, for \ca\ -- see dashed line in
\Fref{fig1} -- we obtain
\beqs\beq \label{resm2}
0.96\lesssim\ns\lesssim0.9654,\>\>\>0.1\lesssim N\lesssim
65,\>\>\>0.05\lesssim{\Dex}/{10^{-2}}\lesssim
16.7\>\>\>\mbox{and}\>\>\> 0.0025\lesssim {r}\lesssim0.039\eeq
with $\wrh\simeq-0.05$ and $\Ns\simeq50$. On the other hand, for
\cb\ -- see dot-dashed line in \Fref{fig1} -- we obtain
\beq \label{resm3} 0.963\lesssim\ns\lesssim0.965,\>\>\>0.1\lesssim
N\lesssim 55,\>\>\>0.23\lesssim{\Dex}/{10^{-2}}\lesssim
8.5\>\>\>\mbox{and}\>\>\> 0.0001\lesssim {r}\lesssim0.04\eeq\eeqs
with $\wrh\simeq(0.25-0.39)$ and $\Ns\simeq54-56$. In both
equations above the lower bound on $N$ is just artificial. For
$N=10$, specific values of parameters and observables are arranged
in the rightmost columns of the Table in \Fref{fig1}.

\section{PI With a Gauge non-Singlet Inflaton}\label{fhi1}

In the present scheme the inflaton field can be identified with
the radial component of a conjugate pair of Higgs superfields. We
here focus on the Higgs superfields, $\bar\Phi$ and $\Phi$, with
$B-L=-1,~1$ which break the GUT symmetry $\Ggut=\Gsm\times
U(1)_{B-L}$ down to SM gauge group $\Gsm$ through their v.e.vs. We
below outline the SUGRA setting in \Sref{fhi10} its inflationary
outcome in \Sref{fhi11}) and its predictions in Sec.~\ref{num1}.
We here update the results of \cref{sor}, taking into account the
recent data of \cref{gws}, and enrich its content adding the model
\hb.

\subsection{SUGRA Set-up}\label{fhi10}

\begin{floatingtable}[r]
\begin{tabular}{|l||lll|}\hline
{\sc Superfields}&$S$&$\Phi$&$\bar\Phi$\\\hline\hline
$U(1)_{B-L}$&$0$&$1$&$-1$\\\hline
$R$ &$1$&$0$&$0$\\\hline
\end{tabular}
\caption {\sl Charge assignments of the superfields.}\label{ch}
\end{floatingtable}

In accordance with the imposed symmetries -- see \Tref{ch} -- we
here adopt the following $W$ -- cf.~\cref{jean}:
\beq W= S\lf \frac12\ldb\bar\Phi\Phi+\ldd
(\bar\Phi\Phi)^2-\frac14M^2\rg,\label{whih} \eeq
where $\ldb, \ldd$ and $M$ are free parameters. In contrast to
\Eref{whi}, we here include the first allowed non-renormalizable
term.  As we see below, this term assist us to activate the
pole-elimination method for \dhi\ and generates a \hb. On the
other hand, the invariance of $K$ under $\Ggut$ enforces us to
introduce a pole of order two within the kinetic terms of
$\phcb-\phc$ system. One possible option -- for another equivalent
one see \cref{sor} -- is
\beq\kba=-N\ln\left(1-|\phc|^2-|\phcb|^2\right)\>\>\>\mbox{or}\>\>\>
\tkba=-N\ln\frac{1-|\phc|^2-|\phcb|^2}{(1-2\phcb\phc)^{1/2}(1-2\phcb^*\phc^*)^{1/2}},\label{tkba}\eeq
which parameterizes the manifold ${\cal
M}_{21}=SU(2,1)/(SU(2)\times U(1))$ \cite{sor} with scalar
curvature ${\cal R}_{21}=-{6}/{N}$ -- note that the present $N$ is
twice that defined in the first paper of \cref{sor}. From the
selected above $W$ and $K$'s, the following inflationary models
emerge:

\begin{itemize}

\item \dhi, where we employ
\beqs\beq \kbba=\kb+\kba \label{kbba} \eeq
and ensure an elimination of the singular denominator appearing in
$\Vhi$ setting
\beq N=2~~~\mbox{and}~~~\tss=-\ldd/\ldb\simeq
1+\ts\,~~\mbox{with}~~~\ts\sim0~~~\mbox{and}~~~M\ll1.\label{kbba1}
\eeq\eeqs

\item \ha\ and \hb, which do not display any singularity in
$\Vhi$, employing \beq \tkbba=\kb+\tkba \label{tkbba}\eeq with
free parameters $N$, $\ldb$, $\ldd$ and $M$. Their discrimination
depends on which of the two inflaton-dependent terms in
\Eref{whih} dominates -- see below.

\end{itemize}

\subsection{Structure of the Inflationary Potential}\label{fhi11}

As in \Sref{fhi31}, we determine the inflationary potential,
$\Vhi$, selecting a suitable parameterization of the involved
superfields. In particular, we set \beq \Phi=\phi
e^{i\theta}\cos\theta_\Phi\>\>\> \mbox{and}\>\>\>\bar\Phi=\phi
e^{i\thb}\sin\theta_\Phi\>\>\>
\mbox{with}\>\>\>0\leq\thn\leq{\pi}/{2}~~~\mbox{and}~~~S= \lf{s
+i\bar s}\rg/{\sqrt{2}}.\eeq
We can easily verify that a D-flat direction is \beq
\vevi{\theta}=\vevi{\thb}=0,\>\vevi{\thn}={\pi/4}\>\>\>\mbox{and}\>\>\>\vevi{S}=0,\label{inftr}\eeq
which can be qualified as inflationary path. Indeed, for both
$K$'s in \Eref{tkbas}, the D term due to $B-L$ symmetry during PI
is \beq \vevi{{\rm D}_{BL}}=
N\lf|\vevi{\phc}|^2-|\vevi{\phcb}|^2\rg/
\lf1-|\vevi{\phc}|^2-|\vevi{\phcb}|^2\rg=0.\eeq
Also, regarding the exponential prefactor of $V_{\rm F}$ in
\Eref{Vsugra} we obtain
\beq \label{eKh} \vevi{e^{K}}=\begin{cases}
\frr^{-N}&\mbox{for}\>\>\>K=\kba,\\
1& \mbox{for}\>\>\>K=\tkba,\end{cases}\eeq
Substituting it and \eqs{K2}{whih} into \Eref{Vhio}, this takes
its master form
\beq\Vhi=\frac{\ld^2}{16} \begin{cases}
{\lf  \sg^2-\tss\sg^4-\mmb^2\rg^2}/{\frr^{N}}&\mbox{for \dhi},\\
{\lf \sg^2-\tss\sg^4-\mmb^2\rg^2}&\mbox{for \ha},\\
{\lf \sg^4-\tts\sg^2-\mmc^2\rg^2}&\mbox{for
\hb},\end{cases}\label{Vmabh}\eeq
where $r_{ij}=-\ld_i/\ld_j$ with $i,j=1,2$ and $\ld$ and $M_i$ are
identified as follows
\beq\ld= \begin{cases}
\ldb~~\mbox{and}~~\mmb=M/\sqrt{\ldb}&\mbox{for~\dhi\ and \ha},\\
\ldd~~\mbox{and}~~\mmc=M/\sqrt{\ldd}&\mbox{for~\hb}.\end{cases}\label{ldbd}\eeq
From \Eref{Vmabh}, we infer that  the pole in $\frr$ is presumably
present in $\Vhi$ of \dhi\ but it disappears in $\Vhi$ of \ha\ and
\hb\ and so no $N$ dependence in $\Vhi$ arises. The elimination of
the pole in the regime of \Eref{kbba1} lets open the realization
of \dhi, though.

\renewcommand{\arraystretch}{1.4}
\begin{table}[!t]
\begin{center}
\begin{tabular}{|c||c|c||l|l|}\hline {\sc
Fields}&{\sc Eigen-}& \multicolumn{3}{|c|}{\sc Masses
Squared}\\\cline{3-5} &{\sc states}& &
\multicolumn{1}{|c|}{$K=\kbba$}&\multicolumn{1}{|c|}{$K=\tkbba$}\\
\hline\hline
%
2 real&$\widehat\theta_{+}$&$m_{\widehat\theta+}^2$&
\multicolumn{2}{|c|}{$3\Hhi^2$}\\\cline{4-5}
scalars&$\widehat \theta_\Phi$ &$\widehat m_{
\theta_\Phi}^2$&\multicolumn{2}{|c|}{$M^2_{BL}+6\Hhi^2(1+2/N-2/N\sg^2)$}
\\\hline
1 complex&$s, {\bar{s}}$ &$ \widehat m_{
s}^2$&\multicolumn{1}{|c|}{$6\Hhi^2(1/N_S-8(1-\sg^2)/N+N\sg^2/2$}&\multicolumn{1}{|c|}{$6\Hhi^2(1/N_S-4/N$}\\
scalar&&&\multicolumn{1}{|c|}{$+2(1-2\sg^2)+8\sg^2/N)$}&\multicolumn{1}{|c|}{$+2/N\sg^2+2\sg^2/N)$}\\\hline
1 gauge boson &{$A_{BL}$}&{$M_{BL}^2$}&
\multicolumn{2}{|c|}{$2Ng^2\sg^2/\frr^2$}\\\hline
$4$ Weyl  & $\what \psi_\pm$ & $\what m^2_{ \psi\pm}$&
\multicolumn{2}{|c|}{${12\frr^2\Hhi^2/N^2\sg^2}$}\\\cline{4-5}
spinors&$\ldu_{BL}, \widehat\psi_{\Phi-}$&$M_{BL}^2$&\multicolumn{2}{|c|}{$2Ng^2\sg^2/\frr^2$}\\
\hline
\end{tabular}\end{center}
\caption{\sl\ftn Mass spectrum the models of HI along the
inflationary trajectory of Eq.~(4.8). }\label{tab1}
\end{table}\renewcommand{\arraystretch}{1.}

To obtain PI we have to correctly identify the canonically
normalized (hatted) fields of the $\phcb-\phc$ system, defined as
follows
\beqs\beq \vevi{K_{\al\bbet}}\dot Z^\al \dot Z^{*\bbet} \simeq
\frac12\lf\dot{\widehat \sg}^2+\dot{\widehat
\th}_+^2+\dot{\widehat \th}_-^2+\dot{\widehat
\th}_\Phi^2\rg~~~\mbox{for}~~~\al=2,3. \label{kzzn}\eeq
-- recall that $Z^1=S$ is already canonically normalized for
$\vevi{S}=0$ as in \Eref{inftr}. We find
\beq \lf \vevi{K_{\al\bbet}}\rg=
\vevi{M_{\phc\phcb}}~~\mbox{with}~~
\vevi{M_{\phc\phcb}}=\frac{\kp\sg^2}{2}\mtta{2/\sg^2-1}{1}{1}{2/\sg^2-1},
\>\> \kp=\frac{N}{\frr^{2}}.\eeq\eeqs
We then diagonalize  $\vevi{M_{\phc\phcb}}$ via a similarity
transformation, i.e.,
\beq U_{\phc\phcb} \vevi{M_{\phc\phcb}} U_{\phc\phcb}^\tr
=\diag\lf \kp_+,\kp_-\rg,\>\>\>\mbox{where}\>\>\>U_{\phc\phcb}=
\frac{1}{\sqrt{2}}\mtt{1}{1}{-1}{1},\>\>
\kp_+=\kp\>\>\>\mbox{and}\>\>\> \kp_-=\kp\frr\,. \eeq
Inserting the expressions above in \Eref{kzzn} we obtain the
hatted fields
\beq \frac{d\se}{d\sg}=J={\sqrt{2N}\over\frr},~~\widehat{\theta}_+
\simeq\sqrt{\kp_+}\sg\theta_+,~~\widehat{\theta}_-
\simeq\sqrt{{\kp_-}}\sg\theta_-~~\mbox{and}~~~\widehat \theta_\Phi
\simeq \sg\sqrt{2\kp_-}\lf\theta_\Phi-{\pi}/{4}\rg,\eeq
where $\th_{\pm}=\lf\bar\th\pm\th\rg/\sqrt{2}$. From the first
equation above we conclude that \Eref{VJe} for $p=2$ is reproduced
for $N_2=2N$. We expect that \dhi\ has similar behavior with \dci,
found in \Sref{fhi31} whereas \ha\ [\hb] may be interpreted as
supersymmetrization of the non-SUSY models with $p=2$ in
\Eref{action1} and $n=4$ [$n=8$] in \Eref{vn}.

Having defined the canonically normalized scalar fields, we can
derive the mass spectrum of our models along the direction of
\Eref{inftr} and verify its stability against the fluctuations of
the non-inflaton fields. Approximate, quite precise though,
expressions for $\sg=\sgx\sim1$ are arranged in \Tref{tab1}. We
confine ourselves to the limits $\ts=\mmb=0$ for \dhi,
$\tss=\mmb=0$ for \ha\ and $\tts=\mmc=0$ for \hb. As in the case
of the spectrum in \Tref{tab3}, $N_S<6$ plays a crucial role in
retaining positive and heavy enough $\what m^2_{s}$. Here,
however, we also display  the masses, $M_{BL}$, of the gauge boson
$A_{BL}$ (which signals the fact that $U(1)_{B-L}$ is broken
during PI) and the masses of the corresponding fermions. The
unspecified eigenstate $\what \psi_\pm$ is defined as \beq \what
\psi_\pm=(\what{\psi}_{\Phi+}\pm
{\psi}_{S})/\sqrt{2}\>\>\>\mbox{where}\>\>\>\psi_{\Phi\pm}=(\psi_\Phi\pm\psi_{\bar\Phi})/\sqrt{2}\,,\eeq
with the spinors $\psi_S$ and $\psi_{\Phi\pm}$ being associated
with the superfields $S$ and $\bar\Phi-\Phi$. It is also evident
that $A_{BL}$ becomes massive absorbing the massless Goldstone
boson associated with $\what\th_-$.

The breakdown of $U(1)_{B-L}$ during PI is crucial in order to
avoid the production of topological defects during the $B-L$ phase
transition, which takes place after end of PI. Indeed, along the
direction of \Eref{inftr}, $\Vhi$ develops a SUSY vacuum lying at
the direction
\beq \label{vev}\vev{S}=0~~~\mbox{and}~~~ %
\vev{\sg}= \begin{cases}
\lf 1-(1-4\tss\mmb^2)^{1/2}\rg^{1/2}/\sqrt{2\tss}&\mbox{for~\dhi\ and \ha},\\
\lf
\tts+(\tts^2+4\mmc^2)^{1/2}\rg^{1/2}/\sqrt{2}&\mbox{for~\hb},\end{cases}\eeq
i.e., $U(1)_{B-L}$ is finally spontaneously broken via the v.e.v
of $\sg$.

\subsection{Results}\label{num1}

As in \Sref{num3}, we here focus on our numerical results -- our
analytic ones for \dhi\ and \ha\ are presented in \cref{sor}.
After enforcing \eqs{Nhi}{Prob} -- which yield $\ld$ together with
$\sgx$ -- the free parameters of the models
$$\mbox{\dhi, ~\ha, ~\hb\ ~~are}~~~(\ts,\mmb), (N, \tss, \mmb)
~~\mbox{and}~~ (N,\tts,\mmc), $$
respectively. Recall that we use $N=2$ exclusively for \dhi. Also,
we determine $\mmb$ and $\mmc$ demanding that the GUT scale within
MSSM $\Mgut\simeq2/2.433\times10^{-2}$ is identified with the
value of $M_{BL}$ -- see \Tref{tab1} -- at the vacuum of
\Eref{vev}, I.e.,
\beq \label{Mg} \vev{M_{BL}}={\sqrt{2N}g\vev{\sg}\over
\vev{f_2}}=\mgut~~\Rightarrow~~\vev{\sg}\simeq\frac{\mgut}{g\sqrt{2N}}
~~\mbox{with}~~~g\simeq0.7,~~\vev{f_2}\simeq1\eeq
and $\vev{\sg}$ given by \Eref{vev}. By varying the remaining
parameters for each model we obtain the allowed curves in the
$\ns-r$ plane-- see \Fref{fig2}. A comparison with the
observational data is also displayed there.  We observe that:

\begin{figure}[t]\vspace*{-0.6cm}
\begin{minipage}{75mm}
\includegraphics[height=9cm,angle=-90]{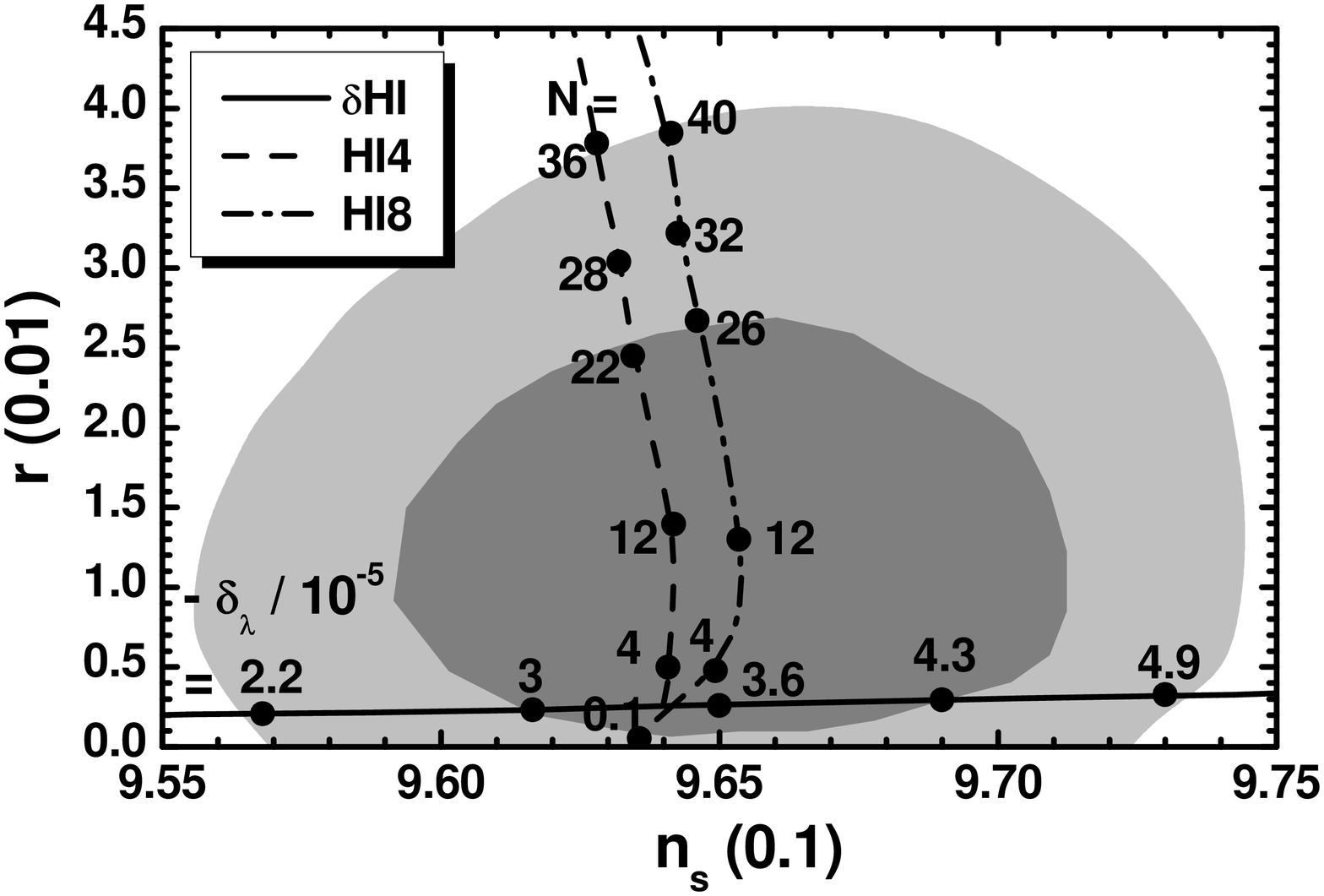}
\end{minipage}
\hfil
\begin{minipage}{95mm}
\begin{center}
{\small
\begin{tabular}{|c||c|c|c|}\hline
{Model:} &{\dhi}&{\ha}&{\hb}\\\hline
${\ts}$ / $\tss$ / $\tts$&$-3.6\cdot10^{-5}$&{$0.01$}&{$10^{-6}$}\\
$N$&{$2$}&$12$&$12$\\\hline\hline
$\sgx/0.1$&$9.9555$&$9.75$&{$9.877$}\\
$\Dex (\%)$&$0.445$&$2.5$&{$1.23$}\\
$\sgf/0.1$&{$5.9$}&$3.9$&{$6.5$}\\\hline
$\wrh$&$0.33$&$0.266$&{$0.58$}\\
$\Ns$&$55.2$&$56.4$&{$58$}\\\hline
$\ld/10^{-5}$&$3.6$&$8.6$&{$8.5$}\\\hline
$\ns/0.1$&$9.65$&$9.64$&{$9.65$}\\
$-\as/10^{-4}$&$6.6$ &$6.4$&{$5.98$}\\
$r/10^{-2}$&$0.26$&$1.4$&{$1.3$}\\\hline
\end{tabular}}
\end{center}
\end{minipage}
\caption{\sl Allowed curves in the $\ns-r$ plane fixing
$\mbl=\mgut$ for (i) \textsf{\ftn $\delta$HI} and various $\ts$'s
indicated on the solid line or (ii) \textsf{\ftn HI4} and
$\tss=0.01$ or \textsf{\ftn HI8} and $\tts=10^{-6}$ and various
$N$'s indicated on the dashed and dot-dashed line respectively.
The shaded corridors are identified as in Fig.~1. The relevant
field values, parameters and observables corresponding to points
shown in the plot are listed in the Table.}
\label{fig2}\end{figure}

\subparagraph{\bf (a)} For \dhi\ -- see the solid line in
\Fref{fig2} -- we obtain results similar to those obtained for
\dci\ in \Sref{num3}. Namely, the resulting $\ns$ and $r$ increase
with $|\ts|$ with $\ns$ covering the whole allowed range in
\Eref{nswmap}. From the considered data we collect the results
\beq \label{resh1}2\lesssim-\ts/10^{-5}\lesssim5.5, \>\>\>
2\lesssim
{r/10^{-3}}\lesssim3.6\>\>\>\mbox{and}\>\>\>4\lesssim\Dex/10^{-3}\lesssim4.75.
\eeq
Also, we obtain $\Ns\simeq(54.8-55.7)$ consistently with
\Eref{Nhi} and the resulting $\wrh\simeq0.3$ from \Eref{wrh}.
Fixing $\ns=0.965$ we find $\ts=-3.6\cdot 10^{-5}$ and $r=0.0026$
-- see the leftmost column of the Table in \Fref{fig2}. \Eref{Mg}
gives $\mmb=0.00587$.

\subparagraph{\bf (b)}  For \ha\ and \hb, $\ns$ and $r$ increase
with $N$ and $\Dex$ which is larger than that obtained in \dhi.
Namely, $\ns$ approaches its central observational value in
\Eref{nswmap} whereas the bound on $r$ yields an upper bound on
$N$.  More specifically, for \ha\  -- see dashed line in
\Fref{fig2} -- we obtain
\beqs\beq \label{resh2}
0.963\lesssim\ns\lesssim0.964,\>\>\>0.1\lesssim N\lesssim
36,\>\>\>0.09\lesssim{\Dex}/{10^{-2}}\lesssim
7.6\>\>\>\mbox{and}\>\>\> 0.0005\lesssim {r}\lesssim0.039\,,\eeq
with $\wrh\simeq0.3$ and $\Ns\simeq56$. \Eref{Mg} dictates
$\mmb\simeq(0.0013-0.0045)$. On the other hand, for \hb\ -- see
dot-dashed line in \Fref{fig2} -- we obtain
\beq \label{resh3} 0.963\lesssim\ns\lesssim0.965,\>\>\>0.1\lesssim
N\lesssim40,\>\>\>0.45\lesssim{\Dex}/{10^{-2}}\lesssim
3.8\>\>\>\mbox{and}\>\>\> 0.0001\lesssim
{r}\lesssim0.039\,,\eeq\eeqs
with $\wrh\simeq(0.25-0.6)$ and $\Ns\simeq(54.6-60)$. \Eref{Mg}
implies $\mmc\simeq(1.1-690)\cdot10^{-6}$. In both equations above
the lower bound on $N$ is just artificial -- as in
\eqs{resm2}{resm3}. For $N=12$, specific values of parameters and
observables are arranged in the rightmost columns of the Table in
\Fref{fig2}.  Although \hb\ is worse than \ha\  regarding the
tuning of $\mmc$ and $\tts$, it leads to $\ns$ values precisely
equal to its central observational one -- cf. \Eref{nswmap}.


\section{Conclusions}\label{con}

We reviewed the implementation of PI first in a non-SUSY and then
to a SUSY framework. In the former regime, we confined ourselves
to models displaying a kinetic mixing in the inflaton sector with
a pole of order one or two and verified their agreement with
observations. In the latter regime, we presented two classes of
models (CI and HI) depending on whether the inflaton is included
into a gauge singlet or non-singlet field. CI and HI are relied on
the superpotential in \eqs{whi}{whih} respectively which respects
an $R$ symmetry and include an inflaton accompanying field which
facilitates the establishment of PI. In each class of models we
singled out three subclasses of models (\dci, \ca\ and \cb) and
(\dhi, \ha\ and \hb). The models \dci\ and \dhi\ are based on the
\Ka s in \eqs{kbas}{kbba} whereas (\ca, \cb) and (\ha, \hb) in
those shown in \eqs{tkbas}{tkbba}. All those \Ka s parameterize
hyperbolic internal geometries with a kinetic pole of order one
for CI and two for HI. The Higgflaton in the last case implements
the breaking of a gauge $U(1)_{B-L}$ symmetry at a scale which may
assume a value compatible with the MSSM unification.

All the models excellently match the observations by restricting
the free parameters to reasonably ample regions of values. In
particular, within \dci\ and \dhi\ any observationally acceptable
$\ns$ is attainable by tuning $\rs$ and $\ts$ respectively to
values of the order $10^{-5}$, whereas $r$ is kept at the level of
$10^{-3}$ -- see \eqs{resm1}{resh1}. On the other hand, \ca, \cb,
\ha\ and \hb\ avoid any tuning, larger $r$'s are achievable as $N$
increases beyond $2$, while $\ns$ lies close to its central
observational value -- see \eqs{resm2}{resm3} for CI and
\eqs{resh2}{resh3} for HI.

\paragraph*{\small \bf\scshape Acknowledgments} {\small  I would like to thank  A.~Marrani,
S.~Ketov and E.W. Kolb for interesting discussions. This research
work was supported by the Hellenic Foundation for Research and
Innovation (H.F.R.I.) under the ``First Call for H.F.R.I. Research
Projects to support Faculty members and Researchers and the
procurement of high-cost research equipment grant'' (Project
Number: 2251).}

\def\ijmp#1#2#3{{\emph{Int. Jour. Mod. Phys.}}
{\bf #1},~#3~(#2)}
\def\plb#1#2#3{{\emph{Phys. Lett.  B }}{\bf #1},~#3~(#2)}
\def\prl#1#2#3{{\emph{Phys. Rev. Lett.} }
{\bf #1},~#3~(#2)}
\def\prep#1#2#3{\emph{Phys. Rep. }{\bf #1},~#3~(#2)}
\def\prd#1#2#3{{\emph{Phys. Rev. } D }{\bf #1},~#3~(#2)}
\def\npb#1#2#3{{\emph{Nucl. Phys.} }{\bf B#1},~#3~(#2)}
\def\astp#1#2#3{\emph{Astropart. Phys.}
{\bf #1}, #3 (#2)}
\def\epjc#1#2#3{{Eur. Phys. J. C}
{\bf #1},~#3~(#2)}
\def\jhep#1#2#3{{\emph{JHEP} }
{\bf #1}, #3 (#2)}
\def\jcap#1#2#3{{\emph{JCAP} }
{\bf #1}, #3 (#2)}
\def\jcapn#1#2#3#4{{\emph{JCAP} }{\bf #1}, no.\ #4, #3 (#2)}

\end{document}